\documentclass[preprint,12pt]{elsarticle}

\usepackage{graphicx}
\usepackage{amsmath}
\usepackage{amssymb}
\usepackage{mathtools}
\usepackage[frozencache,cachedir=minted-cache]{minted}
\usepackage{xcolor}
\definecolor{nicered}{rgb}{0.5,0.,0.}
\definecolor{nicegreen}{rgb}{0.,0.5,0.}
\definecolor{niceblue}{rgb}{0.,0.,0.5}
\usepackage{hyperref}
\hypersetup{colorlinks,citecolor=nicegreen,linkcolor=nicered,urlcolor=niceblue}
\usepackage{siunitx}
\usepackage{xparse}
\usepackage{bm}
\usepackage{soul}
\usepackage{import}
\usepackage{url}
\usepackage{booktabs}

\ExplSyntaxOn
\NewDocumentCommand{\comm}{mm}
{
  \str_case:nnF {#1}
  {
    {info}{\textcolor{blue}{#2}}
    {todo}{\textcolor{red}{#2}}
  }
  {#2}
}
\ExplSyntaxOff

\newcommand{\vv}[1]{\bm{\mathbf{#1}}}
\newcommand{\ddp}{\partial}

\newcommand{\as}{a_s}

\newcommand{\fsub}{\widetilde f}
\newcommand{\df}{\delta f}
\newcommand{\ot}{\otimes}
\newcommand{\dd}{\mathrm{d}}
\newcommand{\FC}{\mathrm{FC}}
\newcommand{\FE}{\mathrm{FE}}
\newcommand{\Sub}{\mathrm{Sub}}
\newcommand{\GMVFN}{\mathrm{GMVFN}}
\newcommand{\NLO}{\mathrm{NLO}}

\biboptions{sort&compress}

\newcounter{bla}

\journal{Computer Physics Communications}

\begin{document}

\begin{frontmatter}



\title{\texttt{Candia-v2}: Logarithmic expansions for DGLAP evolution in $x$-space}


\author[a]{Casey Hampson\corref{author}}
\author[a]{Marco Guzzi\corref{author}}

\cortext[author] {Corresponding authors.\\\textit{E-mail address:} champso1@students.kennesaw.edu, mguzzi@kennesaw.edu}
\address[a]{Department of Physics, Kennesaw State University, Kennesaw, GA 30144, USA}

\begin{abstract}

We present \texttt{Candia-v2}, an open-source software package that generalizes the $x$-space \texttt{Candia} algorithm to next-to-next-to-next-to-leading order (N$^{3}$LO) accuracy in Quantum Chromodynamics (QCD). The code solves the DGLAP evolution equations for unpolarized nucleon parton densities using a highly efficient logarithmic expansion technique that can be systematically extended to all orders in QCD\@.
\texttt{Candia-v2} supersedes the previous original \texttt{C} implementation of the algorithm with significantly increased efficiency and improved API\@. The software is publicly available on GitHub under the GPLv3 license.

\pagebreak

\noindent \textbf{PROGRAM SUMMARY}

\begin{small}
\noindent
{\em Program Title:} \texttt{Candia-v2}                       \\
{\em CPC Library link to program files:} (to be added by Technical Editor) \\
{\em Developer's repository link:} https://github.com/champso1/candia-v2 \\
{\em Licensing provisions(please choose one):} GPLv3  \\
{\em Programming language:} \texttt{C++/Fortran}                        \\
{\em Journal reference of previous version:} https://doi.org/10.1016/j.cpc.2008.06.004    \\
{\em Does the new version supersede the previous version?:} Yes \\
{\em Reasons for the new version:} Significant developments in the calculation of the DGLAP splitting functions at four loops as well as the operator matrix elements at N$^3$LO.\\
{\em Summary of revisions:} Conversion from \texttt{C} to \texttt{C++}, optimizations with convolution/interpolation routines, N$^3$LO evolution, and possibility for LHAPDF inputs and outputs. Implementation of a module for subtraction and residual parton distributions functions (PDFs) \\
{\em Nature of problem:} This program solves the DGLAP evolution equations in $x$-space for PDFs at N$^3$LO \\
{\em Solution method:} The algorithm is based upon logarithmic expansions of the solution in $x$-space and a set of recursion relations for unknown coefficients. \\
{\em Additional comments including restrictions and unusual features:} Due to the dependency on the GNU Scientific Library (GSL) and (optionally) LHAPDF, which do not provide native binaries for Windows and are also difficult to build from source on Windows, we recommend using WSL if a UNIX-like system is unavailable. Additionally, on MacOS systems, since Apple \texttt{clang} does not implement the Intel Threading Building Blocks (TBB) library, it is recommended to either use GCC or LLVM's \texttt{clang}.

\end{small}
\end{abstract}
\end{frontmatter}

\pagebreak
\tableofcontents
\pagebreak

\section{Introduction}
\label{sec:intro}
In this manuscript, we present the public release of the \texttt{Candia-v2} evolution package, which solves the Dokshitzer-Gribov-Lipatov-Altarelli-Parisi (DGLAP)~\cite{Gribov:1972rt,Gribov:1972ri,Lipatov:1974qm,Dokshitzer:1977sg,Altarelli:1977zs} integro-differential equations governing the factorization-scale evolution of collinear parton distribution functions (PDFs) of the proton up to next-to-next-to-next-to-leading order (N$^3$LO) in perturbative QCD. 
\texttt{Candia-v2} implements the $x$-space algorithm documented in Refs.~\cite{Hampson:2025pvi,Cafarella:2005zj,Guzzi:2006wx,Cafarella:2008du} based on efficient logarithmic expansions of the DGLAP solution, whose coefficients are determined recursively directly in $x$-space. The algorithm provides insight into the analytical structure of DGLAP solutions which are expressed in terms of power series. The new version of the code is publicly available at \url{https://github.com/champso1/candia-v2}~\cite{candiav2}.

The package incorporates the most recent results on the 4-loop DGLAP splitting functions as well as the 3-loop operator matrix elements (OMEs) that control the N$^3$LO part of the evolution and the heavy-quark matching conditions respectively. In particular, \texttt{Candia-v2} uses the recent exact calculation for the 4-loop non-singlet (NS) splitting functions presented in Ref.~\cite{Gehrmann:2026qbl} (properties and implications are studied in Ref.~\cite{Moch:2026qsw}) while the singlet (S) sector uses the approximations documented in Refs.~\cite{Davies:2022ofz,Moch:2021qrk,Falcioni:2023luc,Falcioni:2023vqq,Moch:2023tdj,Falcioni:2024xyt,Falcioni:2024qpd,Falcioni:2025hfz} (FHMRUVV). The OMEs are evaluated through the fast interface of the \texttt{libome} library~\cite{Ablinger:2025joi}, which provides users with access to the exact results recently derived in Refs.~\cite{Ablinger:2025nnq,Ablinger:2025joi,Ablinger:2024qxg,Ablinger:2024xtt,Ablinger:2023ahe,Bierenbaum:2022biv,Ablinger:2022wbb,Behring:2021asx,Ablinger:2020snj,Ablinger:2019etw,Ablinger:2018brx,Ablinger:2017xml,Ablinger:2017err,Ablinger:2014tla,Ablinger:2014uka,Behring:2014eya,transitionAqg,Ablinger:2012qm,Blumlein:2012vq,Blumlein:2011mi,Ablinger:2010ty,Bierenbaum:2009zt,Bierenbaum:2008yu,Bierenbaum:2007qe,Bierenbaum:2007dm}.   
 
The ongoing Run 3 of the Large Hadron Collider (LHC) is expected to deliver approximately (300~fb$^{-1}$) of integrated luminosity, and the future high-luminosity LHC (HL-LHC) program aims to increase this amount by about one order of magnitude for both the ATLAS and CMS experiments. In parallel, the Electron-Ion Collider (EIC) will provide high-precision measurements of hadronic structure over a complementary kinematic range~\cite{AbdulKhalek:2021gbh}. These experimental advances require theoretical predictions for hadronic cross sections with comparable accuracy. This, in turn, motivates the development of precise and efficient PDF-evolution tools, such as (\texttt{Candia-v2}), capable of implementing the most recent perturbative QCD ingredients.

Accurate parton evolution is an essential ingredient in the theoretical interpretation of present and future collider data. Over the years, a variety of numerical and semi-analytic strategies have been developed to solve the DGLAP equations, including methods based on Mellin-space techniques, recursive solutions, interpolation grids, Runge-Kutta integration, and parton-branching formulations~\cite{Furmanski:1981cw,Rossi:1983xz,DaLuzVieira:1990xk,Kumano:1992vd,Pascaud:1994vx,Miyama:1995bd,Pascaud:1996ci,Kosower:1997hg,Hirai:1997gb,Blumlein:1997em,Coriano:1998wj,Santorelli:1998yt,Ratcliffe:2000kp,Weinzierl:2002mv,Cafarella:2005zj,Jadach:2005bf,Guzzi:2006wx,Jadach:2007qa,Simonelli:2024vyh}. Several of these approaches have been implemented in fast evolution programs, some of which are publicly available. Representative examples include \texttt{partonevolution}~\cite{Weinzierl:2002mv}, based on Mellin inversion along an optimized contour; \texttt{Pegasus}~\cite{Vogt:2004ns}, which uses the $U$-matrix method in Mellin space; \texttt{Candia}~\cite{Cafarella:2008du,candiav1}, which implements logarithmic recursion relations directly in $x$-space; \texttt{Hoppet}~\cite{Salam:2008qg,Karlberg:2026kte}, based on $x$-space Runge-Kutta evolution; and \texttt{QCDNum}~\cite{Botje:2010ay}, which employs polynomial-spline interpolation in $x$-space. More recent developments include \texttt{Apfel}~\cite{Bertone:2013vaa,Bertone:2017gds}, which combines Runge-Kutta evolution with high-order $x$-space interpolation; \linebreak \texttt{ChiliPDF}~\cite{Diehl:2021gvs}, based on global Chebyshev interpolation in Mellin space; \texttt{EKO}~\cite{Candido:2022tld}, which implements the evolution-kernel-operator formalism in Mellin space; and \texttt{uPDFevolv}~\cite{Hautmann:2014uua,Jung:2024uwc}, which follows the parton-branching approach. 
A recent comparison of QCD evolution at approximate N$^3$LO (aN$^3$LO) accuracy was presented in Ref.~\cite{Cooper-Sarkar:2024crx}.
Global PDF analyses at aN$^3$LO in QCD have recently been documented in Refs.~\cite{NNPDF:2024nan,McGowan:2022nag} while the inclusion of QED corrections with a photon PDF at NLO accuracy is documented in Refs.~\cite{Cridge:2023ryv,Ball:2025xgq}.

The manuscript is organized as follows.
In Sec.~\ref{sec:program-description} we briefly illustrate the algorithm for the N$^3$LO evolution and summarize the recursion relations from Ref.~\cite{Hampson:2025pvi}. In Sec.~\ref{sec:results}, we illustrate the \texttt{Candia-v2} evolution results for selected PDF combinations and provide numerical values in Table~\ref{tab:scaleratioN3LO}. Sec.~\ref{sub-ref-PDFs} discusses subtraction and residual PDFs while in Sec.~\ref{sec:prog-desc}, we provide a detailed description of the program and its functionalities with illustration of the core classes.  
Sec.~\ref{sec:conclusions} contains our concluding remarks.

\section{Brief description of the algorithm}
\label{sec:program-description}

In this section we give a short description of the \texttt{Candia} algorithm at N$^3$LO in QCD and set up the notation. For more details we refer the reader to Ref.~\cite{Hampson:2025pvi} and references therein, where all formal aspects of the algorithm are discussed.   

The DGLAP integro-differential equations can be written as
\begin{equation}    
\frac{\partial}{\partial\ln Q^{2}}f_i(x,Q^{2}) = \sum_{j=q,\bar{q},g} {\cal P}_{ij}(x,\alpha_{s}(Q^2)) \otimes f_j(x,Q^{2})
\label{DGLAP}
\end{equation}
where $f_i(x,Q^2)$ is the PDF for parton $i=q,\bar{q}, g$ in the proton computed at the longitudinal momentum fraction $x$ and energy scale $Q^2$. ${\cal P}_{ij}$ represents the DGLAP splitting functions which is a matrix in the singlet sector and a scalar function in the NS sector.  
The convolution product is defined as
\begin{equation}
  \left[a \otimes b\right](x) = \int_{x}^{1}\frac{\dd y}{y}a\left(\frac{x}{y}\right)b(y) = \int_{x}^{1}\frac{\dd y}{y}a(y)b\left(\frac{x}{y}\right)
\end{equation}
and the general perturbative expansion of the splitting functions is 
\begin{equation}
{\mathcal P}(x,\alpha_s(Q^2)) = \sum_{n=0}^{\infty} \left(\frac{\alpha_s(Q^2)}{2\pi}\right)^{n+1} P^{(n)}(x)
\label{kernel-expansion}
\end{equation}
where $P^{(0)}(x)$ was first computed in Refs.~\cite{Gross:1973id,Politzer:1974fr,Altarelli:1977zs}, $P^{(1)}(x)$ in Refs.~\cite{Floratos:1977au,Floratos:1978ny,Gonzalez-Arroyo:1979guc,Gonzalez-Arroyo:1979qht,Curci:1980uw,Furmanski:1980cm,Floratos:1981hs,Hamberg:1991qt}, and $P^{(2)}(x)$ was presented in Refs.~\cite{Moch:2004pa,Vogt:2004mw}. The calculation of $P^{(3)}(x)$ for the NS sector has been recently published in Ref.~\cite{Gehrmann:2026qbl}, and the calculation for the singlet sector is currently in progress~\cite{Gracey:1994nn,Davies:2016jie,Moch:2017uml,Gehrmann:2023cqm,Falcioni:2023tzp,Gehrmann:2023iah,Kniehl:2025ttz}.  

The renormalization scale ($\mu_R$) evolution of the QCD strong coupling is described by the renormalization group equations (RGEs)  
\begin{equation}
\frac{\ddp \alpha_s(\mu_R)}{\ddp \ln \mu_R^2} = \beta(\alpha_s) = -\sum_{n=0}^{\infty} \frac{\beta_n}{(4\pi)^{n+1}} \alpha_s^{n+2},
\label{eq:1-basics-betafn}
\end{equation}
where the $\beta_n$ coefficients of the QCD $\beta$-function~\cite{Gross:1973id,Politzer:1974fr}  are known up to four loops~\cite{vanRitbergen:1997va,Czakon:2004bu} and we report them below for consistency
\begin{align}
\beta_{0}  &= \frac{11}{3}N_{C}-\frac{4}{3}T_{f} \\
\beta_{1}  &=  \frac{34}{3}N_{C}^{2}-\frac{20}{3}N_{C}T_{f}-4C_{F}T_{f}   
\\
\beta_{2}  &= 
\frac{2857}{54}N_{C}^{3}+2C_{F}^{2}T_{f}-\frac{205}{9}C_{F}N_{C}T_{f}-\frac{1415}{27}N_{C}^{2}T_{f} \nonumber\\
&+\frac{44}{9}C_{F}T_{f}^{2}+\frac{158}{27}N_{C}T_{f}^{2} 
\\
\beta_3 &=   \left( \frac{149753}{6} + 3564 \zeta_3 \right)
        - \left( \frac{1078361}{162} + \frac{6508}{27} \zeta_3 \right) n_f
  \nonumber \\ 
       &+ \left( \frac{50065}{162} + \frac{6472}{81} \zeta_3 \right) n_f^2
       +  \frac{1093}{729}  n_f^3\,,
\label{eq:beta3}
\end{align}
where $\beta_3$ is from Ref.~\cite{vanRitbergen:1997va} and the
fundamental color factors (Casimir operators and normalization constants in QCD) are defined as

\begin{equation}
N_{C}=3,\qquad C_{F}=\frac{N_{C}^{2}-1}{2N_{C}}=\frac{4}{3},\qquad
T_{f}=T_{R}n_{f}=\frac{1}{2}n_{f}.
\end{equation}
$N_{C}$ is the number of colors, and $n_{f}$ is the number of active flavors according to the condition $m_{q}\leq \mu_F$ for a given factorization scale
$\mu_F$, where $m_q$ is the mass of the $q$ quark. Note that in $\beta_3$ these coefficients have been already substituted in with $N_C=3$. 

\subsection{The Non-Singlet Sector}
\label{subsec:ns-sector}

After a change of variable, the evolution of the NS sector is represented by the following scalar equations
\begin{equation}
\frac{\partial}{\partial\alpha_s}f_i(x,\alpha_s) = \frac{P_{i}(x,\alpha_{s})}{\beta(\alpha_s)} \otimes f_i(x,\alpha_s)
\label{NS-DGLAP}
\end{equation}
where $\alpha_s(\mu_F^2)\equiv\alpha_s$ is the strong coupling evaluated at the factorization scale $\mu_F$ and index $i$ refers to the three independent non-singlet flavor combinations, namely NS$^{+}$, NS$^{-}$ and $\textrm{NS}^{V}\equiv\textrm{NS}^{S}+\textrm{NS}^{-}$ for the ``valence''. These flavor combinations are obtained from
\begin{equation}
q_{i}^{(\pm)}=q_{i}\pm\overline{q}_{i}\,,
~~~~~~q^{(\pm)}=\sum_{i=1}^{n_{f}}q_{i}^{(\pm)},
\label{Flav-comb}
\end{equation}
from which the NS flavor asymmetries are given as
\begin{equation}
q_{\mathrm{NS},ik}^{(\pm)} = q_i^{(\pm)} -q_k^{(\pm)}.
\end{equation}
%
The $q_{\mathrm{NS}}^{(\pm)}$ and the $q^{(-)}$ PDF combinations independently evolve with the splitting functions below
\begin{align}
P_{\mathrm{NS}}^{(\pm)} &= P_{qq}^V \pm P_{q\bar{q}}^V, \\
P_{\mathrm{NS}}^V &= P_{qq}^V - P_{q\bar{q}}^V + n_f(P_{qq}^S - P_{q\bar{q}}^S) \equiv P_{\mathrm{NS}}^- + P_{\mathrm{NS}}^S.
\label{NS-splittings}
\end{align}

\subsection{The Singlet Sector}
\label{sec:s-sector}
In the singlet sector, the DGLAP equation is a matrix equation that mixes the gluon and quark distributions. This is given by
\begin{equation}
\frac{\ddp}{\ddp \alpha_s} \begin{pmatrix} q^{(+)}(x,\alpha_s) \\ g(x,\alpha_s)\end{pmatrix}
  = \frac{1}{\beta(\alpha_s)}\begin{pmatrix}P_{qq}(x, \alpha_s) & P_{qg}(x, \alpha_s) \\ P_{gq}(x, \alpha_s) & P_{gg}(x, \alpha_s)\end{pmatrix}
  \otimes \begin{pmatrix} q^{(+)}(x,\alpha_s) \\ g(x,\alpha_s)\end{pmatrix}.
\end{equation}
The quark-quark splitting function $P_{qq}$ is written as
\begin{equation}
P_{qq}=P_{NS}^{+}+n_{f}\left(P_{qq}^{S}+P_{q\bar{q}}^{S}\right)\equiv P_{NS}^{+}+P_{ps}\,,
\label{Pqq-expr}
\end{equation}
where $P_{ps}$ denotes the pure-singlet term. Similarly, the singlet quark--gluon and gluon--quark splitting functions are written as $P_{qg}=n_f P_{q_i g}$ and $P_{gq}=P_{gq_i}$, with the flavor-independent relations $P_{q_i g}=P_{\bar q_i g}$ and $P_{gq_i}=P_{g\bar q}$.

\subsection{\texorpdfstring{$N^3LO$}{N3LO} recursion relations for the exact solution (NS Sector)}
\label{sec:ns-exact}

In Mellin-space, convolution products are mapped onto algebraic products ($[a\otimes b](N)=a(N)b(N)$) and DGLAP equations in the NS sector are purely algebraic with exact solutions. Motivated by this, we introduce the following exponential-type power series {\it ansatz} in $x$-space at the generic scale $\mu$:
\begin{equation}
f^{\mathrm{N}^m\mathrm{LO}}(x,\mu^2) =
\prod_{i=0}^m \left[
\left(\sum_{n=0}^\infty \frac{a_i(x)^n}{n!}L_i^n\right)_{\otimes}
\;
\right]
f(x,\mu_0^2),
\label{x-ansatz}
\end{equation}
where the index $m$ corresponds to the expansion order of the splitting functions in Eq.~(\ref{kernel-expansion}) (at N$^3$LO $m=3$), $f(x,\mu_0^2)$ is the PDF at the initial scale $\mu_0$, and the $a_i(x)$ functions are determined through recursion relations in $x$ space. The functions $L_0,\dots, L_3$ are logarithmic contributions containing only $\alpha_s(\mu^2)\equiv\alpha_s$, $\alpha_s(\mu_0^2)\equiv\alpha_0$ and the $\beta_n$ coefficients. These logarithmic functions are inherited from the algebraic solution in Mellin-space and are given by 
\begin{align}
L_0 &= \log \left(\frac{\alpha_s}{\alpha_0}\right), 
\\
L_1 &= \log \left(\frac{\alpha_s^2 + \bar{b}\alpha_s + \bar{c}}{\alpha_0^2 + \bar{b}\alpha_0 + \bar{c}}\right), 
\\
L_2 &= \log \left(\frac{\alpha_s - r_1}{\alpha_0 - r_1}\right). 
\\
L_3 &= \frac{1}{\sqrt{-\bar{b}^2+4\bar{c}}}\arctan\left( \frac{(\alpha_s - \alpha_0)\sqrt{-\bar{b}^2+4\bar{c}}}{2\alpha_s\alpha_0 + \bar{b}(\alpha_s+\alpha_0) + 2\bar{c}} \right),
\end{align}
where we have defined $\bar{b} \equiv -2\mathrm{Re}[r_2] = -2\mathrm{Re}[r_3]$ and $\bar{c} = |r_2|^2 = |r_3|^3$ and $r_1$, $r_2$, and $r_3$ are the non-trivial roots of the 4-loop QCD $\beta$-function appearing in the denominator of Eq.~(\ref{NS-DGLAP}) (see Sec.~III of Ref.~\cite{Hampson:2025pvi} for more details). We report their $n_f$ dependent numerical values in Table~\ref{n3lo-roots} for completeness where $r_1$ is real, and $r_2$ and $r_3$ are complex conjugates of each other. These are computed numerically and are hard-coded into the program.
\begin{table}[ht]
\centering
\begin{tabular}{|c|c|c|}
\hline
$n_f$ & $r_1$ & $r_2 = r_3^*$ \\ \hline
1 & -0.96511 & 0.162930 - 0.95357i \\ \hline
2 & -1.03151 & 0.185237 - 1.02991i \\ \hline
3 & -1.11203 & 0.221422 - 1.13108i \\ \hline
4 & -1.20902 & 0.286759 - 1.27279i \\ \hline
5 & -1.32059 & 0.424770 - 1.48548i \\ \hline
6 & -1.42780 & 0.796497 - 1.81681i \\ \hline
\end{tabular}
\caption{Numerical values of the roots of the cubic polynomial in the denominator of Eq.~(\ref{NS-DGLAP}).}
\label{n3lo-roots}
\end{table}

The ansatz in Eq.~(\ref{x-ansatz}) can be extended to all orders in QCD~\cite{Cafarella:2005zj,Cafarella:2008du,Guzzi:2006wx,Hampson:2025pvi} and at N$^3$LO it can be written in a more compact form as
\begin{equation}
  f^{\mathrm{N}^3\mathrm{LO}}(x,\mu^2) = \sum_{s=0}^\infty\sum_{t=0}^s\sum_{m=0}^t\sum_{n=0}^m \frac{L_0^n L_1^{m-n}L_2^{t-m}L_3^{s-t}}{n!(m-n)!(t-m)!(s-t)!}D^s_{t,m,n}(x),
  \label{eq:full-exact-ansatz}
\end{equation}
where $s=n+m+\ell+k$ and $t=n+m+\ell$, and the coefficients $D_{t,m,n}^{s}(x)$ are defined as
\begin{equation}
D_{t,m,n}^{s}(x)= a_{0,n}(x)\otimes
a_{1,m-n}(x)\otimes a_{2,t-m}(x) \otimes a_{3,s-t}(x) \otimes f(x,\mu_0^2),
\end{equation}
and are determined by solving the four recursion relations at N$^3LO$
\begin{align}
  D^s_{t,m,n} &= Z_{11} \otimes D^{s-1}_{t-1,m-1,n-1}, \label{eq:n3lo-resum-1}\tag{\theequation{a}}\\
  D^s_{t,m,n} &= Z_{21}\otimes D^s_{t,m,n+1} + Z_{22} \otimes D^{s-1}_{t-1,m-1,n}, 
                \label{eq:n3lo-resum-2}\tag{\theequation{b}}\\
  D^s_{t,m,n} &= Z_{31} \otimes D^s_{t,m+1,n+1} + Z_{32} \otimes D^{s-1}_{t-1,m,n}, 
                \label{eq:n3lo-resum-3}\tag{\theequation{c}}\\
  D^s_{t,m,n} &= Z_{41} \otimes D^s_{t+1,m+1,n+1} + Z_{42} \otimes D^s_{t+1,m+1,n} + Z_{43} \otimes D^{s-1}_{t,m,n},\label{eq:n3lo-resum-4}\tag{\theequation{d}}
\end{align}
where the $Z_{ij}$ coefficients are given as
\begin{align}
  Z_{11}(x) &= R_0(x) = -\frac{2}{\beta_0} P^{(0)}(x),  
              \nonumber\\
  Z_{21} &= \frac{1}{2\beta_3\gamma}\left[ 16\pi^2\beta_1 + 4\pi r_1\beta_2 - (\bar{c} + \bar{b} r_1)\beta_3 \right], 
           \nonumber\\
  Z_{22}(x) &= \frac{1}{\beta_3\gamma}\left[32\pi^2P^{(1)}(x) + 16\pi r_1 P^{(2)}(x) - 8(\bar{c}+\bar{b}r_1)P^{(3)} (x)\right], 
              \nonumber\\
  Z_{31} &=\frac{1}{\beta_3\gamma} \left[(-16\pi^2\beta_1 - 4\pi r_1\beta_2 - r_1^2\beta_3)\right], 
           \nonumber\\
  Z_{32}(x) &= \frac{1}{\beta_3\gamma} \left[- 64\pi^2 P^{(1)}(x) - 32\pi r_1 P^{(2)}(x) - 16 r_1^2 P^{(3)}(x)\right], 
              \nonumber\\
  Z_{41} &= -2 \bar{b},  
           \nonumber\\
  Z_{42} &= \frac{1}{\beta_3\gamma}\left[ 32\pi^2(\bar{b}+r_1)\beta_1  -8\pi \bar{c}\beta_2 - 2\bar{c}r_1\beta_3 \right],  
           \nonumber\\
  Z_{43}(x)& = \frac{1}{\beta_3\gamma}\left[ 128\pi^2(\bar{b}+r_1) P^{(1)}(x) - 64\pi \bar{c} P^{(2)}(x) - 32 \bar{c} r_1 P^{(3)}(x) \right]\,.
             \label{Zij-def}
\end{align}

The procedure for determining all $D^s_{t,m,n}$ coefficients for a particular value of the index $s$ is outlined below
\begin{enumerate}
\item All $D^s_{t,m,n}$ with $n\neq0$ are computed using Eq.~(\ref{eq:n3lo-resum-1}),
\item All $D^s_{s,s,0}$ are computed using Eq.~(\ref{eq:n3lo-resum-2}),
\item All $D^s_{s,m,0}$ are computed using Eq.~(\ref{eq:n3lo-resum-3}) where $m \neq s$, with decreasing $m$,
\item All $D^s_{t,m,0}$ are computed using Eq.~(\ref{eq:n3lo-resum-4}) where $t \neq s$, with decreasing $t$ and $m$.
\end{enumerate}

\subsection{Recursion Relations for the Truncated Solution}
\label{sec:recrels-singlet}

Expanding the ${\cal P}(x,\alpha_s)/\beta(\alpha_s)$ ratio in the DGLAP singlet sector at order ${\cal O}(\alpha_s^k)$, one obtains the $k$-order truncated version of the DGLAP equations 

\begin{eqnarray}
\frac{\partial{{\bf f}(x,\alpha_s)}}{\partial\alpha_s}=
\frac{1}{\alpha_s}\left[{\bf R}_0+\alpha_s {\bf R}_1 +\alpha_s^2
{\bf R}_2+\alpha_s^3
{\bf R}_3 + \dots + \alpha_s^k {\bf R}_k \right]\otimes {\bf f}(x,\alpha_s),
\label{N3LOsinglet}
\end{eqnarray}
where the bold-face characters are used for the vector notation and the ${\bf R}_k$ matrix operators are defined order by order as
\begin{align}
{\bf R}_0(x)&=-\frac{2}{\beta_0}{\bf P}^{(0)}(x),\\
{\bf R}_1(x)&=-\frac{1}{\pi}\left(\frac{\beta_1}{4\beta_0}{\bf R}_0(x) +
\frac{{\bf P}^{(1)}(x)}{\beta_0}\right),
\\
{\bf R}_2(x)&=-\frac{1}{\pi}\left(\frac{{\bf P}^{(2)}(x)}{2\pi\beta_0}
+\frac{{\bf R}_1(x) \beta_1}{4\beta_0}+\frac{{\bf R}_0(x) \beta_2}{16\pi\beta_0}\right), 
\\
{\bf R}_3(x) &= -\frac{1}{\pi}\left(\frac{{\bf P}^{(3)}(x)}{4\pi^2\beta_0} +\frac{{\bf R}^{(2)}(x)\beta_1}{4\beta_0} + \frac{{\bf R}^{(1)}(x)\beta_2}{16\pi\beta_0}+\frac{{\bf R}^{(0)}(x)\beta_3}{64\pi^2\beta_0}\right),\\
&\dots\nonumber
\end{align}

The recursion relations for the case of DGLAP truncated solutions are obtained from the general logarithmic ansatz   
in $x$-space given by
\begin{equation}
\vv{f}(x,\alpha_s) = \sum_{n=0}^\infty \left\{ \left[ \sum_{i=0}^\kappa \alpha_s^i \frac{\vv{S}^i_n(x)}{n!} \right] \ln^n \frac{\alpha_s}{\alpha_0} \right\},
\label{log-ansatz}
\end{equation}
where $\kappa$ represents the truncation index that defines the $\alpha_s$ accuracy of the singlet solution or the NS one, in the case of scalar equations. In general, $\kappa\geq k$ and these two indices are independent.     

At N$^3$LO, the recursion relation for the ${\bf S}^{i}_{n}(x)$ coefficients with $i\geq 3$ is given by

\begin{align}
  \vv{S}_{n+1}^i(x) = &-\frac{\beta_1}{4\pi\beta_0}\vv{S}_{n+1}^{i-1}(x)
-\frac{\beta_2}{16\pi^2\beta_0}\vv{S}_{n+1} ^{i-2}(x)
-\frac{\beta_3}{64\pi^3\beta_0}\vv{S}^{i-3}_{n+1}(x) 
\nonumber\\
&-i\vv{S}_n^i(x)-(i-1)\frac{\beta_1}{4\pi\beta_0}\vv{S}_n^{i-1}(x)
-(i-2)\frac{\beta_2}{16\pi^2\beta_0}\vv{S}_n^{i-2}(x) \nonumber\\
&-(i-3)\frac{\beta_3}{64\pi^3\beta_0}\vv{S}_n^{i-3}(x) \nonumber\\
&-\frac{2}{\beta_0}[\vv{P}^{(0)}\otimes\vv{S}_n^i](x)
-\frac{1}{\pi\beta_0}[\vv{P}^{(1)}\otimes\vv{S}_n^{i-1}](x)\nonumber\\
&-\frac{1}{2\pi^2\beta_0}[\vv{P}^{(2)}\otimes\vv{S}_n^{i-2}](x)
-\frac{1}{4\pi^3\beta_0}[\vv{P}^{(3)}\otimes\vv{S}_n^{i-3}](x)\,,
\label{eq:singlet-n3lo}
\end{align}

which needs to be combined with the $i=2$, 1, and 0 pieces corresponding to the NNLO, NLO, and LO relations respectively, from Ref.~\cite{Hampson:2025pvi}.

These recursion relations generate truncated solutions at arbitrary order $\kappa$ that are equivalent those obtained using the $U$-matrix method~\cite{Blumlein:1997em,Vogt:2004ns,Cafarella:2005zj}.  

\section{Results}
\label{sec:results}
The results of \texttt{Candia-v2} evolution at aN$^3$LO in QCD are reported in Table~\ref{tab:scaleratioN3LO} where we also show the impact from scale dependence as discussed in Ref.~\cite{Hampson:2025pvi}. The initial conditions are those from the Les Houches toy model used in Sec.~4.4 of Ref.~\cite{Dittmar:2005ed} at the initial scale of $\mu^2_0=2$ GeV$^2$. The PDFs are evolved in the variable flavor number scheme (VFNS) up to $\mu_F^2 = 10^4$ GeV$^2$. 
The heavy-quark mass values passed in the input are in the pole-mass approximation and are set to $m_c=\mu_0$, $m_b=4.5$ GeV, and $m_t=175$ GeV. The aN$^3$LO splitting functions currently implemented in the code use the approximations documented in Refs.~\cite{Davies:2022ofz,Moch:2021qrk,Falcioni:2023luc,Falcioni:2023vqq,Moch:2023tdj,Falcioni:2024xyt,Falcioni:2024qpd,Falcioni:2025hfz} for the singlet sector, while for the NS sector we implemented the exact results recently calculated in Ref.~\cite{Gehrmann:2026qbl}. At NNLO, we use the approximations in Refs.~\cite{Moch:2004pa,Vogt:2004mw} for a fast evaluation. The 3-loop OMEs for heavy-quark threshold conditions beyond NNLO with single mass from Refs.~\cite{Ablinger:2025joi,Ablinger:2024xtt} are implemented using the  \texttt{libome}~\cite{Ablinger:2025joi,libome} publicly available \texttt{C++} libraries where their numerical representation is given in terms of precise local overlapping series expansion. For the 2-loop OMEs, we implement the analytical expressions of Refs.~\cite{Buza:1995ie,Buza:1996wv}.

\begin{table}[htp]
\scriptsize
\noindent\makebox[\textwidth][c]{
\begin{tabular}{ccccccccc}
\toprule
$x$ & $xuv$ & $xdv$ & $xL_-$ & $xL_+$ & $xs_+$ & $xc_+$ & $xb_+$ & $xg$ \\
\midrule
\multicolumn{9}{c}{$\mu_{\rm r}^2 = 1.0\mu_{\rm f}^2$} \\
\midrule
1e$-5$ &  3.020e$-3$ &  1.750e$-3$ &  1.231e$-4$ &  3.546e$+1$ &  1.706e$+1$ &  1.612e$+1$ &  1.315e$+1$ &  2.224e$+2$ \\
1e$-4$ &  1.408e$-2$ &  8.241e$-3$ &  4.776e$-4$ &  1.561e$+1$ &  7.272e$+0$ &  6.785e$+0$ &  5.329e$+0$ &  8.859e$+1$ \\
1e$-3$ &  6.083e$-2$ &  3.508e$-2$ &  1.752e$-3$ &  6.382e$+0$ &  2.779e$+0$ &  2.520e$+0$ &  1.851e$+0$ &  3.034e$+1$ \\
1e$-2$ &  2.336e$-1$ &  1.307e$-1$ &  5.822e$-3$ &  2.267e$+0$ &  8.542e$-1$ &  7.045e$-1$ &  4.623e$-1$ &  7.786e$+0$ \\
1e$-1$ &  5.485e$-1$ &  2.695e$-1$ &  9.995e$-3$ &  3.845e$-1$ &  1.125e$-1$ &  6.830e$-2$ &  3.790e$-2$ &  8.496e$-1$ \\
3e$-1$ &  3.444e$-1$ &  1.276e$-1$ &  2.946e$-3$ &  3.457e$-2$ &  8.887e$-3$ &  3.966e$-3$ &  2.085e$-3$ &  7.870e$-2$ \\
5e$-1$ &  1.179e$-1$ &  3.060e$-2$ &  3.653e$-4$ &  2.321e$-3$ &  5.681e$-4$ &  2.018e$-4$ &  1.138e$-4$ &  7.634e$-3$ \\
7e$-1$ &  1.933e$-2$ &  2.965e$-3$ &  1.285e$-5$ &  5.243e$-5$ &  1.266e$-5$ &  3.403e$-6$ &  2.496e$-6$ &  3.710e$-4$ \\
9e$-1$ &  3.315e$-4$ &  1.674e$-5$ &  8.112e$-9$ &  2.500e$-8$ &  6.555e$-9$ &  6.732e$-10$ &  1.429e$-9$ &  1.163e$-6$ \\
\midrule
\multicolumn{9}{c}{$\mu_{\rm r}^2 = 0.5\mu_{\rm f}^2$} \\
\midrule
1e$-5$ &  2.865e$-3$ &  1.552e$-3$ &  1.340e$-4$ &  3.692e$+1$ &  1.779e$+1$ &  1.585e$+1$ &  1.342e$+1$ &  2.300e$+2$ \\
1e$-4$ &  1.424e$-2$ &  8.263e$-3$ &  5.135e$-4$ &  1.582e$+1$ &  7.378e$+0$ &  6.665e$+0$ &  5.392e$+0$ &  8.999e$+1$ \\
1e$-3$ &  6.210e$-2$ &  3.589e$-2$ &  1.864e$-3$ &  6.391e$+0$ &  2.786e$+0$ &  2.568e$+0$ &  1.884e$+0$ &  3.064e$+1$ \\
1e$-2$ &  2.358e$-1$ &  1.320e$-1$ &  6.018e$-3$ &  2.260e$+0$ &  8.529e$-1$ &  7.506e$-1$ &  4.711e$-1$ &  7.818e$+0$ \\
1e$-1$ &  5.458e$-1$ &  2.679e$-1$ &  9.979e$-3$ &  3.823e$-1$ &  1.124e$-1$ &  7.316e$-2$ &  3.805e$-2$ &  8.421e$-1$ \\
3e$-1$ &  3.402e$-1$ &  1.259e$-1$ &  2.905e$-3$ &  3.425e$-2$ &  8.873e$-3$ &  3.724e$-3$ &  2.029e$-3$ &  7.749e$-2$ \\
5e$-1$ &  1.159e$-1$ &  3.002e$-2$ &  3.580e$-4$ &  2.290e$-3$ &  5.660e$-4$ &  1.227e$-4$ &  1.050e$-4$ &  7.495e$-3$ \\
7e$-1$ &  1.888e$-2$ &  2.892e$-3$ &  1.251e$-5$ &  5.139e$-5$ &  1.252e$-5$ & -1.207e$-6$ &  2.014e$-6$ &  3.631e$-4$ \\
9e$-1$ &  3.204e$-4$ &  1.614e$-5$ &  7.803e$-9$ &  2.289e$-8$ &  5.728e$-9$ & -6.525e$-9$ &  6.654e$-11$ &  1.118e$-6$ \\
\midrule
\multicolumn{9}{c}{$\mu_{\rm r}^2 = 2.0\mu_{\rm f}^2$} \\
\midrule
1e$-5$ &  3.005e$-3$ &  1.774e$-3$ &  1.136e$-4$ &  3.444e$+1$ &  1.654e$+1$ &  1.589e$+1$ &  1.298e$+1$ &  2.189e$+2$ \\
1e$-4$ &  1.381e$-2$ &  8.100e$-3$ &  4.467e$-4$ &  1.532e$+1$ &  7.128e$+0$ &  6.709e$+0$ &  5.271e$+0$ &  8.768e$+1$ \\
1e$-3$ &  5.989e$-2$ &  3.452e$-2$ &  1.668e$-3$ &  6.310e$+0$ &  2.742e$+0$ &  2.471e$+0$ &  1.827e$+0$ &  3.010e$+1$ \\
1e$-2$ &  2.321e$-1$ &  1.300e$-1$ &  5.691e$-3$ &  2.259e$+0$ &  8.486e$-1$ &  6.830e$-1$ &  4.566e$-1$ &  7.758e$+0$ \\
1e$-1$ &  5.505e$-1$ &  2.708e$-1$ &  1.002e$-2$ &  3.868e$-1$ &  1.128e$-1$ &  6.609e$-2$ &  3.777e$-2$ &  8.548e$-1$ \\
3e$-1$ &  3.478e$-1$ &  1.290e$-1$ &  2.980e$-3$ &  3.499e$-2$ &  8.974e$-3$ &  3.954e$-3$ &  2.105e$-3$ &  7.964e$-2$ \\
5e$-1$ &  1.196e$-1$ &  3.109e$-2$ &  3.717e$-4$ &  2.358e$-3$ &  5.754e$-4$ &  2.164e$-4$ &  1.171e$-4$ &  7.748e$-3$ \\
7e$-1$ &  1.972e$-2$ &  3.030e$-3$ &  1.316e$-5$ &  5.343e$-5$ &  1.284e$-5$ &  4.419e$-6$ &  2.677e$-6$ &  3.776e$-4$ \\
9e$-1$ &  3.419e$-4$ &  1.729e$-5$ &  8.400e$-9$ &  2.566e$-8$ &  6.689e$-9$ &  2.478e$-9$ &  1.977e$-9$ &  1.198e$-6$ \\
\bottomrule
\end{tabular}%
}
\caption{N$^3$LO evolution with factorization/renormalization scale dependence. The initial conditions are from the Les Houches toy model described in Sec. 4.4 of Ref.~\cite{Dittmar:2005ed}.}
\label{tab:scaleratioN3LO}
\end{table}

\begin{figure}
\centering    \includegraphics[width=1.0\linewidth]{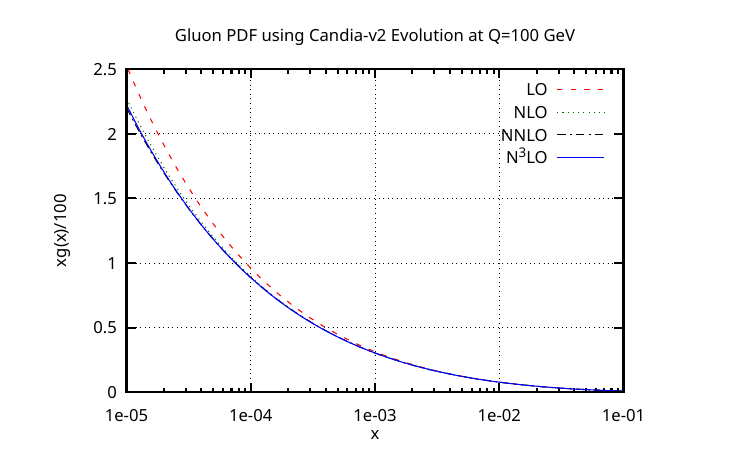}
\includegraphics[width=1.0\linewidth]{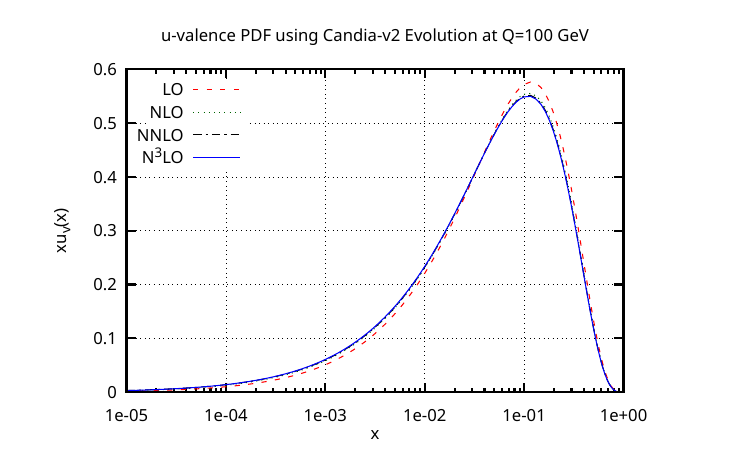}
\caption{Results for the \texttt{Candia-v2} evolution of the gluon and the $u_v$ PDFs. }
\label{g-uv-PDFs}
\end{figure}
\begin{figure}
\centering    \includegraphics[width=1.0\linewidth]{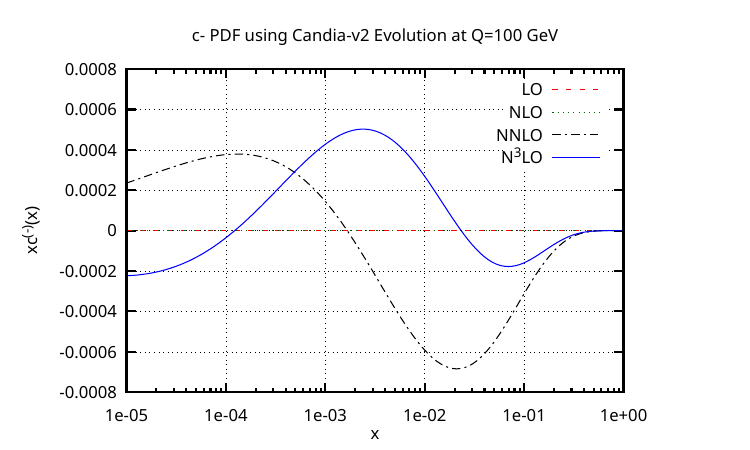}
\includegraphics[width=1.0\linewidth]{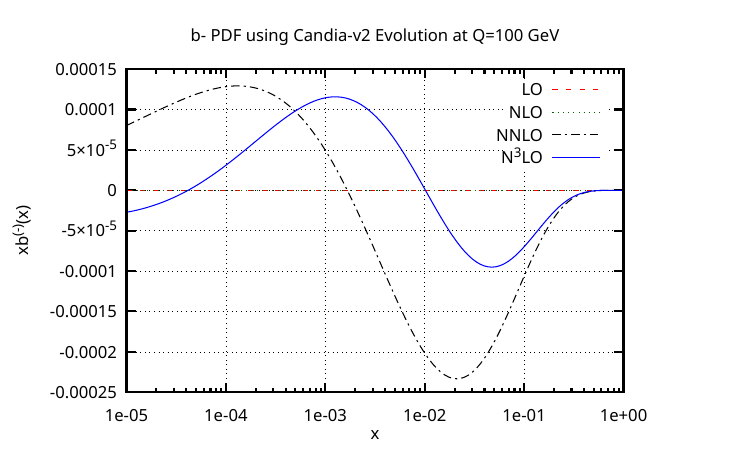}
\caption{Results for the \texttt{Candia-v2} evolution of the $c^{(-)}$ and $b^{(-)}$ heavy flavor combinations.}
\label{cm-bm-PDFs}
\end{figure}

In Fig.~\ref{g-uv-PDFs}, we illustrate the results for the $Q=100$ GeV \texttt{Candia-v2} evolution at the various perturbative orders and the convergence pattern for the gluon and $u_v$ PDFs. The results for the $c^{(-)}(x,Q^2) = c(x,Q^2) - \bar{c}(x,Q^2)$ and $b^{(-)} = b(x,Q^2) - \bar{b}(x,Q^2)$ heavy-flavor combinations are illustrated in Fig.~\ref{cm-bm-PDFs}. These heavy-flavor asymmetries start to be different from zero at NNLO and Fig.~\ref{cm-bm-PDFs} shows the impact from both the evolution and heavy-flavor matching conditions (see Sec.~6 of Ref.~\cite{Hampson:2025pvi}) at N$^3$LO for the charm and bottom PDFs.

\newpage 

\section{Subtraction and Residual PDFs}
\label{sub-ref-PDFs}

In the calculation of hadronic cross sections whose domain is defined over a wide range of energies, General Mass Variable Flavor Number (GMVFN) schemes are factorization schemes that organize heavy-quark
contributions so as to interpolate between the fixed-flavor description near threshold and the massless variable-flavor description at scales much larger than the heavy-quark mass. For example, in a GMVFN scheme such as the ACOT~\cite{Aivazis:1993kh,Aivazis:1993pi,Tung:2001mv,Kramer:2000hn,Guzzi:2011ew,Guzzi:2024can} scheme, the hadronic cross section may be written schematically as
\begin{equation}
  \dd\sigma^{\GMVFN}
  = \dd\sigma^{\FC}+\dd\sigma^{\FE}-\dd\sigma^{\Sub} \, ,
  \label{eq:gmvfn_master}
\end{equation}
where $\dd\sigma^{\FC}$ denotes flavor-creation (FC) contributions in which heavy quarks are produced in the final state and their full mass dependence is retained in the calculation;
$\dd\sigma^{\FE}$ denotes flavor-excitation (FE) contributions that are initiated by a heavy-quark PDF and represent the excitation of a heavy flavor in the proton; and $\dd\sigma^{\Sub}$ is a subtraction (SUB) contribution which removes the overlap between the two in the various kinematic regimes.
This subtraction is required because the collinear limit of the FC channel is already resummed into the heavy-quark PDF $f_Q(x,\mu)$.

As observed in Ref.~\cite{Guzzi:2024can}, the SUB terms can be organized together using universal objects called \emph{subtraction PDFs}. To NLO and NNLO, that is, ${\cal O}(\alpha_s^2)$ and ${\cal O}(\alpha_s^3)$ respectively in the subtraction
coefficients, one may define
\begin{align}
\fsub_Q^{(1)}(x,\mu)
&= \as\,\bigl[A_{Qg}^{(1)} \ot g\bigr](x,\mu),
\label{eq:fsub1}
\\
\fsub_Q^{(2)}(x,\mu)
&= \as^2\sum_{i=g,q,\bar q}
\bigl[A_{Qi}^{(2)} \ot f_i\bigr](x,\mu),
\\
\fsub_Q^{(3)}(x,\mu)
&= \as^3\sum_{i=g, q,\bar q}
\bigl[A_{Qi}^{(3)} \ot f_i\bigr](x,\mu),
\label{eq:fsub2}
\\
\fsub_Q^{(\NLO)}(x,\mu)
&\equiv \fsub_Q^{(1)}(x,\mu)+\fsub_Q^{(2)}(x,\mu).
\label{eq:fsubnlo}
\\
\fsub_Q^{(\textrm{NNLO})}(x,\mu)
&\equiv \fsub_Q^{(1)}(x,\mu)+\fsub_Q^{(2)}(x,\mu)+\fsub_Q^{(3)}(x,\mu),
\label{eq:fsubnnlo}
\end{align}
which refer to the heavy quark $Q$ and where we use $a_s\equiv\alpha_s/(4\pi)$.
The functions $A_{Qi}^{(n)}$ are the  OMEs arising from mass factorization. They describe the perturbative transition of a massless
initial-state parton $i$ into the heavy quark $Q$ and contain the dependence on
$m_Q$ and on the factorization scheme. They are universal in that they are independent of the
particular short-distance hard process. For this reason, the quantities in
Eqs.~\eqref{eq:fsub1}--\eqref{eq:fsubnnlo} are universal PDF-level matching
objects: once computed for a chosen PDF ensemble, heavy-quark mass, perturbative
order, and scheme, they can be tabulated and reused in different implementations.

It is more economical to combine the FE and SUB contributions before performing the hard-scattering convolution because FE and SUB contributions share the same matrix elements. This is
achieved by defining \emph{residual PDFs}
\begin{align}
  \df_Q^{(1)}(x,\mu)
  &\equiv f_Q(x,\mu)-\fsub_Q^{(1)}(x,\mu),
  \label{eq:res1}
  \\
  \df_Q^{(\NLO)}(x,\mu)
  &\equiv f_Q(x,\mu)-\fsub_Q^{(\NLO)}(x,\mu).
  \label{eq:resnlo}
\end{align}
The residual $\df_Q$ is the part of the evolved heavy-quark PDF that is not already accounted for by the fixed-order collinear subtraction through the specified perturbative order.  With these definitions,  
FE and SUB contributions are combined directly using residual
PDFs. This combination aims to facilitate the numerical implementation of GMVFN schemes currently in use \cite{Aivazis:1993kh,Aivazis:1993pi,Tung:2001mv,Kramer:2000hn,Guzzi:2011ew,Guzzi:2024can,Buza:1996wv,Thorne:1997ga,Thorne:2006qt,Cacciari:1998it,Forte:2010ta,Bonvini:2015pxa,Bonvini:2016fgf}. The same subtraction or residual PDF grids can be convoluted with the FE contributions where different approximations in the treatment of mass effects are allowed and correspond to different scheme choices.

\begin{figure}
\centering
\includegraphics[width=1.0\linewidth]{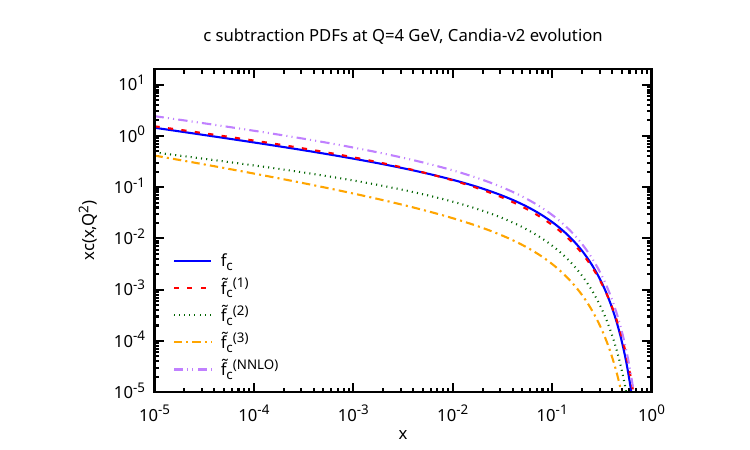}
\includegraphics[width=1.0\linewidth]{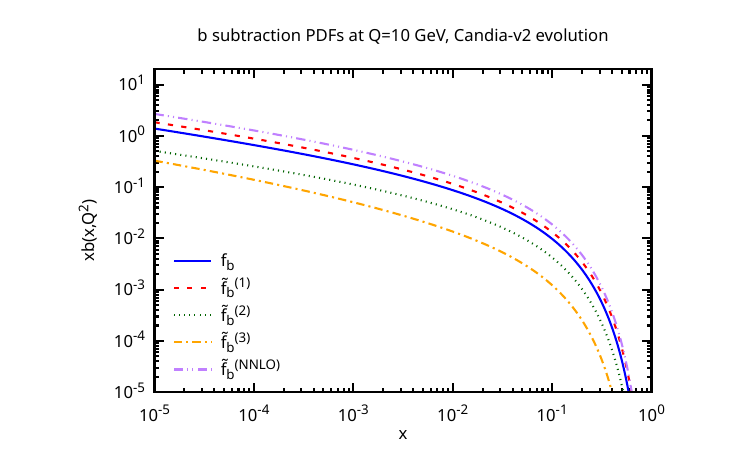}
\caption{Subtraction PDFs for the charm and bottom quark evolved at $Q=4$ GeV and $Q=10$ GeV respectively.}
\label{sub-PDFs}
\end{figure}

In Fig.~\ref{sub-PDFs}, we show the charm- and bottom-quark subtraction PDFs evolved with \texttt{Candia-v2} at $Q=4$ GeV ($n_f=4$) and $Q=10$ GeV ($n_f=5$) respectively, which are compared to the charm and bottom PDFs from the Les Houches toy model of Ref.~\cite{Dittmar:2005ed} evolved at the same energies and $n_f$'s. The generation of subtraction and residual PDFs in \texttt{Candia-v2} is very simple and the procedure is outlined in detail in Sec.~\ref{subsec:subpdf-access}.

\newpage

\section{Description of the Program}
\label{sec:prog-desc}

In this section we describe the requirements to acquire, build, and use \texttt{Candia-v2}, followed by discussions on the different parts of the library. 

\subsection{Requirements and Dependencies}
\label{subsec:reqs-and-deps}

There are some version requirements and dependencies one needs to install in order to successfully compile and run the code. They are:
\begin{itemize}
\item \texttt{CMake} and a build tool e.g. \texttt{GNU Make} or \texttt{Ninja},
\item \texttt{C++20} compatibility -- among the standard compilers, one would need \texttt{GCC >= 13} or \texttt{LLVM CLang >= 17} (see Section~\ref{subsubsec:compilers} for a note on the compiler choice),
\item \texttt{GFortran},
\item \texttt{Intel Threading Building Blocks (TBB)}: the backend for the implementation of \texttt{C++} execution policies,
\item \texttt{GSL} (GNU Scientific Library),
\item (Optional) \LaTeX~utilities: if one wants to build the manuscripts included in the repository,
\item (Optional) \texttt{Doxygen}: if one wants to build the code documentation,
\item (Optional) \texttt{LHAPDF}: if one wants to interface with LHAPDF.
\end{itemize}

The optional dependencies are enabled by passing \texttt{CMake} flags when configuring the project; see Section~\ref{subsec:acq-and-build} for more details.

\subsubsection{A Note on System and Compiler Choices}
\label{subsubsec:compilers}

The two mainstream platform-independent compiler suites, namely \texttt{GCC} and \texttt{LLVM CLang}, have been tested for compiling and using \texttt{Candia-v2} on Linux machines, MacOS machines, and Windows machines inside of Windows Subsystem for Linux (\texttt{WSL}).
The Apple \texttt{CLang} compiler's standard library implementation does not natively include support for \texttt{C++} execution policies, so \texttt{GCC} or \texttt{LLVM CLang} are recommended on MacOS.
Windows is not supported natively nor recommended due to GSL and \texttt{LHAPDF} being autotools-based, as well as the lack of a \texttt{Fortran} compiler within \texttt{Visual Studio}.
Therefore, if using a Windows machine, we highly recommend using \texttt{WSL}.

\subsection{Acquiring and Building}
\label{subsec:acq-and-build}

The code is publicly available at \href{GitHub}{https://github.com/champso1/candia-v2}.
There are several active branches in which various new features and/or fixes are being addressed, but by default the \texttt{main} branch contains the most stable version, and the \texttt{dev} branch contains the most recent version.
While one can at any time build from the \texttt{main} branch, it is preferable to checkout the corresponding tag from the cloned repository via \mintinline{bash}{git checkout <tag>}, e.g., \mintinline{bash}{git checkout v1.7.0}.
After acquiring the source code, one follows the standard \texttt{CMake} building process:
\begin{minted}{bash}
  git clone https://github.com/champso1/candia-v2 \
    --recurse-submodules
  cd candia-v2
  mkdir build && cd build
  cmake .. [-G <Generator>] [<other options>]
  cmake --build .
  cmake --install .
\end{minted}
Note that \mintinline{bash}{--recurse-submodules} is passed to \mintinline{bash}{git clone} so that the \texttt{libome} repository, a submodule, is cloned and initialized correctly.
Among standard \texttt{CMake} options, there are several additional options that are specific to \texttt{Candia-v2}:
\begin{itemize}
\item \mintinline{cfg}{CANDIA_WITH_LHAPDF}\ (default=\texttt{false}): Compile in the LHAPDF interface,
\item \mintinline{cfg}{CANDIA_LHAPDF_DIR}\ (default=\path{/usr/local}): The LHAPDF installation prefix. Should only be specified if \mintinline{cfg}{CANDIA_WITH_LHAPDF} is enabled and the user has installed LHAPDF in a non-standard location, i.e. anywhere other than \path{/usr/local},
\item \mintinline{cfg}{CANDIA_BUILD_MANUSCRIPTS}\ (default=\texttt{false}): Compile the manuscripts from the \path{/manuscripts} directory. Requires \LaTeX~utilities to be available on the command line,
\item \mintinline{cfg}{CANDIA_BUILD_DOCS}\ (default=\texttt{false}): Build the code documentation using \texttt{Doxygen},
\item \mintinline{cfg}{CANDIA_BUILD_EXAMPLES}\ (default=\texttt{true}): Build the two example programs from the \path{examples/} directory, which illustrate the various components of the library,
\item \mintinline{cfg}{CANDIA_BUILD_TESTS}\ (default=\texttt{false}): Build the ``tests''. These are not standard tests, rather they are a large set of additional \texttt{C++} files to perform benchmarking, plot generation, etc.
\end{itemize}

\subsection{Basic User Guide}
\label{subsec:ex-prog}

In this section we present a complete example program that illustrates the capabilities of \texttt{Candia-v2}. Notice that throughout the following code snippets, 
we have defined an alias for an unsigned integer type called \mintinline{cpp}{uint}:

\begin{minted}{cpp}
using uint = unsigned int;
\end{minted}

Listing~\ref{lst:ex-prog} contains a full \texttt{C++} program that performs the N$^3$LO DGLAP evolution at the energy scale $\mu=100$ GeV using the Les Houches standard benchmark initial conditions~\cite{Dittmar:2005ed}.
\begin{listing}
\begin{minted}{cpp}
#include <Candia-v2/Candia.hpp>
using namespace Candia2;

#include <vector>

int main()
{
  const double order = 3;
  const double iterations = 10;
  const double trunc_idx = 10;
  const double Qf = 100;
  const double mur2_muf2 = 1.0;

  std::vector<double> xtab{
    1e-5, 1e-4, 1e-3, 1e-2, 0.1, 0.3, 0.5, 0.7, 0.9};
  Grid grid(xtab);

  LesHouchesDistribution dist(Qf);
  AlphaS alphas(
    order,
    dist.Q0(), dist.Qf(),
    dist.alpha0(),
    mur2_muf2);
  alphas.setVFNS(dist.masses(), dist.nfi(), dist.nff());
  // alphas.setFFNS(4);

  DGLAPSolver solver(
    order, grid, alphas,
    Qf, iterations, trunc_idx,
    dist, mur2_muf2);
  std::vector<ArrayGrid> dists = solver.evolve();
  // do stuff with the distributions
}
\end{minted}
  \caption{Example program to perform an N$^3$LO PDF evolution at $\mu=100$ GeV, using the current Les Houches standard benchmark initial conditions~\cite{Dittmar:2005ed}.}
  \label{lst:ex-prog}
\end{listing}
In the code, the user first includes the main header \mintinline{cpp}{Candia-v2/Candia.hpp},  which includes all the necessary headers so that nothing else needs to be included.
The following variables are also defined:
\begin{itemize}
\item \mintinline{cpp}{order}: the perturbative order of the solution -- this is the $m$ index in N$^m$LO,
\item \mintinline{cpp}{iterations}: the number of total iterations to complete -- this number corresponds to the $s$ index in Eq.~\eqref{eq:full-exact-ansatz} in the case of the exact ansatz, and the $n$ index in Eq.~\eqref{log-ansatz} in the case of the truncated ansatz,
\item \mintinline{cpp}{trunc_idx}: the truncation index -- this corresponds to the variable $\kappa$ in Eq.~\eqref{log-ansatz},
\item \mintinline{cpp}{Qf}: the final evolution energy,
\item \mintinline{cpp}{mur2_muf2}: the ratio $\mu_R^2/\mu_F^2$.
\end{itemize}
Using these variables, the user initializes the main classes handling the interpolation and convolution grid, the running coupling, the initial conditions, and the solver.
An \mintinline{cpp}{std::vector<ArrayGrid>} is returned, which can be processed as required; the \mintinline{cpp}{ArrayGrid} is a simple storage container for accessing array values whose points are determined via a grid (see Section~\ref{subsec:arraygrid} for a lower-level description of this class).
Elements can be accessed like an ordinary \mintinline{cpp}{std::vector}, e.g., using the \mintinline{cpp}{operator[]} syntax. There are 37 total distributions, whose indices are described in Table~\ref{tab:dist-indices}. 

The example file \path{examples/evolve_dglap.cpp} is equivalent to that in Listing~\ref{lst:ex-prog}, but with comments describing each component.

\begin{table}[h]
  \centering
  \begin{tabular}{c|c}
    \hline 
    0 & $g$ \\ 
    1-6 & $q_{i}$ \\ 
    7-12 & $\overline{q}_i$ \\ 
    13-18 & $q_{i}^{(-)}$ \\ 
    19-24 & $q_{i}^{(+)}$ \\ 
    25 & $q^{(-)}$ \\ 
    26-30 & $q_{NS,1i}^{(-)},\quad i\neq1$ \\ 
    31 & $q^{(+)}$ \\ 
    32-36 & $q_{NS,1i}^{(+)},\quad i\neq1$ \\ 
    \hline
  \end{tabular}
  \caption{Correspondence between indices and parton distributions functions.}
  \label{tab:dist-indices}
\end{table}

\subsection{Code Description}
\label{sec:code-description}

This section illustrates a lower-level description of the various classes composing \texttt{Candia-v2}.

\subsubsection{The ArrayGrid Class}
\label{subsec:arraygrid}

The \mintinline{cpp}{ArrayGrid} class is the underlying array type used for PDFs 
or any quantity defined on the $x$-grid, which is described in the next section. The motivation for creating a new class, rather than utilizing nested \mintinline{cpp}{std::vector} 
instances, is to avoid the memory fragmentation issues associated with large numbers of nested arrays 
required for higher-order coefficients. \mintinline{cpp}{ArrayGrid} is an alias for the base 
class \mintinline{cpp}{ArrayGridBase<type, D>}, where \mintinline{cpp}{type} represents the 
underlying data type and \mintinline{cpp}{D} denotes the number of dimensions:

\begin{minted}{cpp}
using ArrayGrid = ArrayGridBase<double, 1>;
\end{minted}

All memory is allocated only once in a single array. Array access is wrapped in the \mintinline{cpp}{operator()} method -- as multi-argument \mintinline{cpp}{operator[]} 
syntax is restricted and is not able to take multiple arguments in \texttt{C++20} -- which handles the dimension strides correctly. For $D=1$, e.g., in the \mintinline{cpp}{ArrayGrid}, a specialization is provided that enables the traditional \mintinline{cpp}{operator[]} syntax like an ordinary array.

The creation of an \mintinline{cpp}{ArrayGridBase} is done by:

\begin{minted}{cpp}
template <typename T, uint D>
ArrayGridBase::ArrayGridBase(
  std::array<uint, D> const& sizes
);
\end{minted}

where \mintinline{cpp}{sizes} is an array of length $D$ which contains the size of each dimension.

Accessing elements in the array is done by calling

\begin{minted}{cpp}
template <typename... TArgs>
decltype(auto) operator()(TArgs... args)
\end{minted}

This method supports either $D-1$ arguments, in which case a \mintinline{cpp}{std::span} is returned (representing the innermost array dimension) or $D$ arguments, in which a single number is returned. Other higher-order views over more dimensions are not supported because they are not used in this code.

\subsubsection{The Grid Class}
\label{subsec:grid-class}

The $x$-grid where the PDFs are defined and on which interpolations and convolutions are done is handled by the \mintinline{cpp}{Grid} class. By default, the grid is filled in three separate segments, which is done in order to best handle the sensitive behavior of PDFs and splitting functions in the extreme $x \rightarrow 0$ and $x \rightarrow 1$ kinematic regions. By default, there are 100 points in the first segment, 50 in the second, and 25 in the last, which can be changed by passing the corresponding arguments to the constructor.

The constructor has the following signature:

\begin{minted}{cpp}
Grid::Grid(
    std::vector<double> const& xtab,
    GridFillerArgs const& grid_filler={},
    ConvIntArgs const& gauleg_args={}
);
\end{minted}

where \mintinline{cpp}{GridFillerArgs} is a structure (struct) containing options for how the grid is filled:

\begin{minted}{cpp}
struct GridFillerArgs final
{
  double min{1.0e-5};
  uint log_size{100};
  uint lin_size{50};
  uint quad_size{25};
  double pivot1{0.1}, pivot2{0.9};
};
\end{minted}

and \mintinline{cpp}{ConvIntArgs} is a struct containing options for convolutions and interpolations:

\begin{minted}{cpp}
struct ConvIntArgs final
{
  uint num_gauss_points{50};
  uint num_interp_points{4};
};
\end{minted}

The \mintinline{cpp}{xtab} array contains specific $x$-values that the user wants aligned with the grid; their corresponding indices are stored in the \mintinline{cpp}{ntab} array (which can be retrieved by calling \mintinline{cpp}{grid.ntab()}) to ensure that values of other quantities at these nodes can be efficiently extracted (e.g., for benchmarking purposes).

The grid class acts as an iterable object, enabling range-based loops as demonstrated in the following snippet (assuming the variable \mintinline{cpp}{grid} is initialized):

\begin{minted}{cpp}
for (double x : grid) {
  // do something with each x
}
\end{minted}

Furthermore, the \mintinline{cpp}{.enumerate()} method permits simultaneous retrieval of both the index and its corresponding $x$-value. This method mirrors the design and functionality of Python's built-in \mintinline{python}{enumerate()} function, and can be used like:

\begin{minted}{cpp}
for (auto [i, x] : grid.enumerate()) {
  // do something with each index i and value x
}
\end{minted}

Finally, the signatures for the convolution and interpolation routines are

\begin{minted}{cpp}
double Grid::interpolate(
  std::span<double> y,
  double x);
double Grid::convolution(
  std::span<double> A,
  Expression& E,
  uint k);
double Grid::convolution(
  std::span<double> A1,
  std::span<double> A2,
  double tau);
\end{minted}

\texttt{Candia-v2} supports two primary convolution modalities: 
convolutions between an array (e.g., a PDF), represented by the \mintinline{cpp}{std::span}, and an expression (e.g., a splitting function), represented by the \mintinline{cpp}{Expression},
and convolutions between two separate arrays, such as those required for parton luminosity calculations.

These convolutions are evaluated numerically via Gauss-Legendre quadrature. 
The corresponding abscissae and weights are precomputed within the interval 
$(x_{\min}, 1)$ during object construction. In a standard integration loop, 
$z$ denotes a single abscissa and $w$ represents its associated weight. 
The underlying convolution relies on a sequence of coordinate mappings 
tailored to each grid segment (logarithmic, linear, or quadratic). 
For a given target $x$-value and the current quadrature point $z$, 
the code evaluates the mapped value $y$ and the corresponding 
Jacobian $J$. This mapping logic is implemented internally via an array 
of \mintinline{cpp}{std::function} objects, managed by a wrapper 
function that dispatches the appropriate transformation for the specified $x$-value.

\subsubsection{The AlphaS Class}
\label{subsec:alphas-class}

The evolution of the QCD strong coupling $\alpha_s$ is handed by the \mintinline{cpp}{AlphaS} class, whose constructor has the following signature:

\begin{minted}{cpp}
AlphaS::AlphaS(
  uint order,
  double Q0, double Qf,
  double alpha0,
  double mur2_muf2
);
\end{minted}

The initial evolution scale $Q_0$ and $\alpha_0\equiv \alpha_s(Q_0)$ are specified by the initial conditions of the evolution, while the perturbative order, the final evolution scale $Q_f$, and the ratio $\mu_R^2/\mu_F^2$, are chosen by the user.

In general, the evolution proceeds via either the 
Fixed-Flavor Number Scheme (FFNS) or the Variable-Flavor Number Scheme (VFNS). 
In the FFNS approach, the number of active flavors remains constant. 
Conversely, in the VFNS framework, the system transitions to $n_f+1$ 
active flavors once the evolution scale crosses the threshold of 
a heavy-quark pole mass. These operational modes are selected by 
invoking either of the following functions:

\begin{minted}{cpp}
void AlphaS::setFFNS(uint nf);
void AlphaS::setVFNS(
  std::array<double, 8> const& masses,
  uint nfi, uint nff);
\end{minted}

The strong coupling values in both schemes are 
stored in two internal arrays, \mintinline{cpp}{pre} and 
\mintinline{cpp}{post}, which store the values of $\alpha_s$ 
immediately before and after a given mass threshold, respectively, 
indexed by the number of active flavors $n_f$. Consequently, 
for a single evolution step within a threshold, 
the initial value of $\alpha_s$ is provided by \mintinline{cpp}{post[nf]}
and the final value is given by \mintinline{cpp}{pre[nf+1]}.
In the FFNS, these represent the only two occupied elements in 
the arrays, as the evolution consists of a single step. 
Intermediate values 
between thresholds are computed by solving the QCD renormalization 
group equation (RGE) via a fourth-order Runge-Kutta algorithm, while values across thresholds are obtained by performing matching in accordance to the perturbative order.
At leading order (LO), the exact analytical solution to the RGE is exploited instead of using the numerical Runge-Kutta solution to optimize computational performance.

This class also handles the calculation of the QCD $\beta$-function and its coefficients:

\begin{minted}{cpp}
double AlphaS::betaFn(double alpha) const;
double AlphaS::beta0() const;
double AlphaS::beta1() const;
double AlphaS::beta2() const;
double AlphaS::beta3() const;
\end{minted}

The methods related to the $\beta$-function coefficients simply retrieve values which are calculated each time the number of flavors $n_f$ is updated via

\begin{minted}{cpp}
void AlphaS::update(uint nf);
\end{minted}

\subsubsection{Distributions}
\label{subsec:distributions}

All initial conditions are handled by the \mintinline{cpp}{Distribution} class and its derived classes. For instance, \mintinline{cpp}{LesHouchesDistribution} is one such class that implements the benchmark initial conditions from Ref.~\cite{Dittmar:2005ed}. These classes define the values of the PDFs and the strong coupling at the initial evolution energy $\mu_0^2 = \qty{2}{\GeV\squared}$, as well as the quark masses and the initial number of flavors. Each subclass must implement the functions that return the initial PDFs for the gluon, as well as $xu(x, Q_0^2)$, $xd(x, Q_0^2)$, and $xs(x, Q_0^2)$ along with the anti-quark distributions. They must also specify the following methods:

\begin{minted}{cpp}
void Distribution::fillSingletCoeffs(
  accessor_type const& accessor,
  std::vector<value_type> const& grid_points
);
void Distribution::fillNonSingletCoeffs(
  accessor_type const& accessor,
  std::vector<value_type> const& grid_points
);
void Distribution::setup(double Q0, double Qf);
\end{minted}

The first two methods compute the initialization of the evolution coefficients using the initial PDFs. The \mintinline{cpp}{accessor} is a function that accepts an ID and a grid index. The ID corresponds to a specific distribution in Table~\ref{tab:dist-indices}, while the grid index specifies the target kinematic point. The \mintinline{cpp}{setup} method defines the actions required when a user updates the initial or final energies, such as returning a new value for the initial $n_f$.

\subsubsection{Expressions}
\label{subsec:expressions}

The main \mintinline{cpp}{Expression} class handles splitting functions 
and operator matrix elements, which are defined as mathematical 
distributions used to regularize and isolate divergences. These 
functions contain regular parts, plus-distributions, and 
$\delta$-function parts. Additionally, due to the method by which the 
convolutions are performed, only the $n_f$-dependent and $x$-independent 
components of the plus-distributions and $\delta$-functions are 
required. Consequently, each expression pre-computes and stores these 
values by implementing:

\begin{minted}{cpp}
void preCalc();
\end{minted}
The value of $n_f$ and any other $x$-independent quantities for the splitting functions are updated globally by calling

\begin{minted}{cpp}
static SplittingFunction::update(
  uint nf,
  double beta0,
  double log_muf2_mur2
);
\end{minted}

The value of any piece of an expression is obtained by

\begin{minted}{cpp}
double Expression::calcRegular(double x);
double Expression::calcPlus();
double Expression::calcDelta();
\end{minted}

while the pre-calculated values of the latter two are obtained by 

\begin{minted}{cpp}
double Expression::plus();
double Expression::delta();
\end{minted}

Every splitting function is a derived class of \mintinline{cpp}{SplittingFunction}, 
which itself inherits from \mintinline{cpp}{Expression}. Note that, by default, the global 
value of $n_f$ is set to 4; therefore, evaluating these functions independently 
requires invoking the static \mintinline{cpp}{update()} method with the desired 
value of $n_f$ and other $x$-independent quantities.

A similar architecture is used for the OMEs. At NNLO, 
each OME is implemented as a derived class of \mintinline{cpp}{OpMatElem}. At 
N$^3$LO, where the package interfaces with \texttt{libome}, a single class 
named \mintinline{cpp}{OpMatElemN3LO} is used. Its constructor accepts an 
\mintinline{cpp}{ome::rpd_distribution}, which serves as a standard 
container for OMEs within \texttt{libome}.

\subsubsection{The DGLAPSolver Class}
\label{subsec:dglapsolver}

The \mintinline{cpp}{DGLAPSolver} class manages the evolution of the PDFs. 
Specifically, it computes all ansatz coefficients utilizing the recursion 
relations defined at the specified perturbative order. In the singlet 
sector, these coefficients are evaluated via the truncated ansatz. For 
the non-singlet sector, the class can compute the coefficients using 
either the truncated or the exact ansatz, depending on the configuration 
specified by the user.

The class constructor has the following signature:

\begin{minted}{cpp}
DGLAPSolver(
  uint order, Grid& grid, AlphaS const& alpha_s,
  double Qf, uint iterations, uint trunc_idx,
  Distribution const& initial_dist,
  double mur2_muf2 = 1.0
);
\end{minted}

Several configuration options can be supplied to the \mintinline{cpp}{DGLAPSolver} 
instance prior to performing the evolution. Chief among these are the approximation 
types for the three-loop splitting functions, $P^{(3)}$. The framework currently supports 
two distinct envelope approximations, alongside a third option corresponding to their 
average. For the non-singlet sector, the exact analytic expressions are also available.

By default, the solver utilizes the average of the envelopes for all singlet splitting 
functions, and the exact expressions for the non-singlet splitting functions.
The \mintinline{cpp}{.setP3ApproximationTypes()} method is used to customize this behavior. For example, to 
select the first envelope for $P^{(3)}_{qg}$ while retaining the default average for the 
remaining singlet functions, and to enforce the envelope average for $P_{\mathrm{NS}}^{(3),+}$
as opposed to the exact expression, the configuration would be specified as follows:
\begin{minted}{cpp}
solver.setP3ApproximationTypes({
    std::make_pair(ExprName::P3qg, P3ApproxType::Imod1),
    std::make_pair(ExprName::P3nsp, P3ApproxType::ImodAvg),
});
\end{minted}

An additional set of options, primarily intended for benchmarking 
and compatibility verification, can be accessed via the \mintinline{cpp}{.getOptions()} 
method. This method returns the configuration structure detailed in Listing~\ref{lst:dglap-options}. 
These parameters permit the selective activation of various N$^3$LO components
or the usage of the Fortran Application Binary Interface (ABI) for evaluating the splitting 
functions.

Once the solver is configured, the evolution is executed by invoking either 
the \mintinline{cpp}{evolve()} or the \mintinline{cpp}{evolveTrunc()} method. The 
former solves the non-singlet sector evolution using the exact ansatz, whereas 
the latter employs the truncated ansatz. At higher perturbative orders, 
the non-singlet coefficients (such as the $D^{s}_{t,m,n}$ coefficients at N$^3$LO) 
acquire additional dimensions. Consequently, the algorithmic time complexity of 
the exact ansatz scales exponentially with the perturbative order, while that of 
the truncated ansatz remains highly manageable. At N$^3$LO, this complexity difference 
becomes pronounced; for large numbers of evolution iterations, the truncated ansatz 
offers a significant performance advantage.

\begin{listing}
\begin{minted}{cpp}
struct DGLAPOptions final
{
  bool use_nnlo_matching_conditions_at_n3lo{false};
  bool disable_heavy_flavor_matching{false};
  bool use_n3lo_heavyquark_asymmetry{true};
  bool use_fortran_nnlo_splitfuncs{false};
  bool use_fortran_n3lo_splitfuncs{false};
};
\end{minted}
  \caption{The \mintinline{cpp}{DGLAPOptions} struct, containing options related to the evolution.}
  \label{lst:dglap-options}
\end{listing}

\subsubsection{LHAPDF Interface}
\label{subsec:lhapdf-interface}

If the corresponding compilation flags are enabled during the \texttt{CMake} 
configuration phase, the code provides an interface to import input 
distributions from, or export evolution results to, the \texttt{LHAPDF} library. 
To handle input initialization, the initial state can be populated directly from 
an \texttt{LHAPDF} set using the \mintinline{cpp}{LHAPDFDistribution} class. This 
class encapsulates an \texttt{LHAPDF} object alongside the user-specified initial 
and final evolution energies. Because \mintinline{cpp}{LHAPDFDistribution} inherits 
from the \mintinline{cpp}{Distribution} base class, it can be utilized in an 
identical manner to the \mintinline{cpp}{LesHouchesDistribution} shown in 
Listing~\ref{lst:ex-prog}.

Conversely, exporting the evolved PDFs into a format 
compatible with \texttt{LHAPDF} is managed by the 
\mintinline{cpp}{LHAPDFGrid} class. This component handles the generation of a standard 
grid output file that can be natively loaded by \texttt{LHAPDF}. A representative 
example demonstrating this is provided in Listing~\ref{lst:lhapdfgrid}, and an example file with additional comments is given in the repository as \path{examples/lhapdf_grid.cpp}.

\begin{listing}
\begin{minted}{cpp}
#include "Candia-v2/Candia.hpp"
#include "Candia-v2/LHAPDFGrid.hpp"
using namespace Candia2;

int main()
{
  std::vector<double> qvals{10.0, 50.0, 100.0};
  LesHouchesDistribution dist(qvals.back());
  Grid grid({
    1e-5, 1e-4, 1e-3, 1e-2, 0.1, 0.3, 0.5, 0.7, 0.9});
  uint order = 3;
  uint iterations = 10;
  uint trunc_idx = 10;
  double mur2_muf2 = 1.0;
  
  LHAPDFGrid lhapdfgrid(
    "testpdf", "infofile.in",
    dist, grid,
    order, iterations, trunc_idx, mur2_muf2);
  // lhapdfgrid.evolve(qvals, {});   // exact
  lhapdfgrid.evolveTrunc(qvals, {}); // truncated
  lhapdfgrid.write();
}
\end{minted}
  \caption{An example program that will create an LHAPDF-formatted grid called ``testpdf''.}
  \label{lst:lhapdfgrid}
\end{listing}

The template file \path{infofile.in} serves as the blueprint for the \path{.info} metadata file mandatory for all \texttt{LHAPDF} PDF sets. An example configuration file is provided in the project root directory; it contains placeholder fields enclosed in percent signs (\%), so that upon successful completion of the PDF evolution, the code substitutes these placeholders with the computed physical specifications and parameters. For instance, the strong coupling $\alpha_s$ values are inserted by replacing the placeholder string \texttt{\%AS\_QS\%} within the file.

Note that \texttt{Candia-v2} does not directly install the output PDFs into the \texttt{LHAPDF} PDF set directory, due to the fact that \texttt{LHAPDF} may be installed in a system path, e.g., \path{/usr/local}. This prevents potential write-permission conflicts or the need for superuser escalation during installation. Instead, generated PDF sets are written to the local working directory. Users can subsequently configure \texttt{LHAPDF} to load the sets locally, or manually transfer the directories to the \texttt{LHAPDF} set directory, typically located at \path{<lhapdf-prefix>/share/LHAPDF}.

\subsubsection{Subtraction and Residual PDFs}
\label{subsec:subpdf-access}

The generation of subtraction and residual PDFs is straightforward, requiring only a single call to the following function:

\begin{minted}{cpp}
std::vector<ArrayGrid> DGLAPSolver::calculateSubtractionPDFs();
\end{minted}

This function returns the same datatype as the \mintinline{cpp}{.evolve()} methods and contains in total 20 PDF objects: $\tilde{f}_q^{(1)}$, $\tilde{f}_q^{(2)}$, $\tilde{f}_q^{(3)}$, $\tilde{f}_q^{(\mathrm{NLO})}$, $\tilde{f}_q^{(\mathrm{NNLO})}$, ${\delta}f_q^{(1)}$, ${\delta}f_q^{(2)}$, ${\delta}f_q^{(3)}$, ${\delta}f_q^{(\mathrm{NLO})}$, and ${\delta}f_q^{(\mathrm{NNLO})}$, where the index $q$ represents either the charm- or the bottom-quark PDF.

Handling the order of these in an array of length 20 is aided by the \mintinline{cpp}{SubtractionPDFIndices} enum\footnote{This is a large enum, so we refer the reader to the source code for the full list of values.} so that, for instance, one can retrieve ${\delta}f_b^{(2)}$ simply using 

\begin{minted}{cpp}
auto deltafb2 = subpdfs[B_DELTAF2];
\end{minted}

If the solver has not evolved up to a scale higher than the mass of the charm or the bottom quark, then those PDFs are zero.

\subsection{Major Fixes and Optimizations from \texorpdfstring{\texttt{Candia-v1}}{Candia-v1}}
\label{subsec:optims}

In this section, we describe the major bug fixes, algorithmic optimizations, and architectural enhancements introduced in \texttt{Candia-v2}. A primary improvement stems from the enhanced flexibility that \texttt{C++} offers over \texttt{C} regarding type-generic functionality and interfaces. Although this transition does not directly improve performance, it significantly aids debugging and improves code readability leading to a faster developing process and cleaner API.

Several bugs present in the original \texttt{Candia} code have been fixed in \texttt{Candia-v2}. First, the NNLO coefficient associated with the exact analytical solution omitted a factor of $\beta_1$ within the arctangent argument. Second, the variable representing the logarithmic ratio $\ln(\mu_R^2/\mu_F^2)$ was incorrectly initialized, leading to its value being set to zero (corresponding to a ratio $\mu_R^2/\mu_F^2 = 1$) regardless of user inputs. This led to an invalid matching procedure for the strong coupling if the user specified a ratio $\mu_R^2/\mu_F^2 \neq 1$.
 
Third, the initial conditions for the coupling evolution were misconfigured. The intent of the Les Houches benchmarking initial conditions is to initiate the evolution at $n_f=3$, immediately \textit{below} the charm-quark mass threshold, in order to trigger matching instantaneously for both the coupling and the PDFs. However, the initial reference value $\alpha_0=0.35$ was inadvertently assigned \textit{above} the charm-quark  threshold, systematically altering the evolved values of $\alpha_s(Q^2)$. Finally, a minor typographical error in the definition of $A_{gg}^{(2)}$ was corrected, though its numerical impact on the overall evolution was found to be negligible.

In addition to critical bug fixes, several improvements have been implemented in \texttt{Candia-v2}. The most significant development is  the implementation of parallel computing to distribute the computation of the evolution equations for the decoupled non-singlet PDFs across independent execution threads. In the singlet sector, DGLAP equations are coupled and the CPU-intensive numerical computations from the convolutions are refactored to allow the utilization of \texttt{C++} execution policies within the \mintinline{cpp}{<algorithms>} library. This allows the compiler to optimize loop iterations dynamically via parallelization, vectorization, or a combination of the two.

The second major enhancement involves optimized memory management. Based on the recursive structure of the algorithm, the code only retains data from adjacent iterations in memory. This drastically reduces the total memory usage, especially at higher perturbative orders. Finally, the specific formulation of the numerical convolution eliminates the need to repeatedly evaluate the $x$-dependent components of the $P_2$ and $P_3$ kernel terms, which correspond to the coefficients of the plus distribution and the Dirac delta function, respectively. Consequently, these terms are computed only once per active flavor number $n_f$ and are subsequently cached, accelerating the evaluation of the splitting functions.

\section{Conclusions}
\label{sec:conclusions}

We have presented \texttt{Candia-v2}, a modernized and extended implementation
of the \texttt{Candia} algorithm for solving the DGLAP evolution equations
directly in Bjorken-$x$ space. The new version supersedes the original
\texttt{Candia} code by providing a \texttt{C++} interface, improved memory management,
parallelized components, and a more flexible API.

The main physics extension of \texttt{Candia-v2} is the implementation of PDF
evolution through approximate N$^3$LO accuracy in QCD. The code includes the currently
available four-loop DGLAP splitting functions, using exact results in the
non-singlet sector and state-of-the-art approximations in the singlet sector.
Heavy-quark threshold matching is implemented with operator matrix elements up
to three loops, interfaced through \texttt{libome}. 

In addition to standard PDF evolution, \texttt{Candia-v2} provides direct access
to subtraction and residual heavy-quark PDFs. These quantities provide reusable, process-independent building blocks that facilitate the implementation of a large variety of GMVFN schemes with improved numerical accuracy.

The package also includes optional interfaces to \texttt{LHAPDF}, enabling both LHAPDF input distributions and LHAPDF-formatted output grids. This makes
\texttt{Candia-v2} suitable both as a standalone evolution code and as a
component in larger environments for broader phenomenological studies. 

The public release of \texttt{Candia-v2} provides a useful tool for analyses in high-precision collider phenomenology. As exact
four-loop singlet splitting functions and additional higher-order ingredients
become available, the modular structure of the code will allow them to be
incorporated in a straightforward way. \texttt{Candia-v2} therefore provides a
flexible platform for precision phenomenology as increasingly precise measurements from the LHC, the HL-LHC program, and future facilities such as the Electron-Ion Collider demand commensurate theoretical accuracy.

\section*{Acknowledgements}
 
M.G. and C.H. are partially supported by the National Science Foundation under Grant No.~PHY-2412071. 
This work used the high-performance computing resource from the Kennesaw State University HPC clusters.

\bibliographystyle{elsarticle-num}

\end{document}